\documentclass[prl,twocolumn,showpacs]{revtex4}
\usepackage{amssymb}
\usepackage{graphicx}
\newcommand{\rmi}{\mathrm{i}}
\newcommand{\rmd}{\,\mathrm{d}}
\renewcommand{\Im}{\mathop{\mathrm{Im}}\nolimits}

\begin{document}
\title{Threshold and high-frequency behavior of dipole-bound anion photodetachment}
\author{V.~E.~Chernov and Boris~A.~Zon}
\email{zon@niif.vsu.ru}
\affiliation{Voronezh State University,%
394693, Russia, Voronezh, University Sq.,1 }

\begin{abstract}
An explicit analytic description is given for dipole-bound anion
(DBA) as an excess electron bound to the molecular neutral due
to its dipole moment. The calculated DBA photodetachment
cross-section displays $\propto\omega^{-2}$ behavior for
large~$\omega$, in complete accordance with the experimental data [Bailey
\textit{et al.}, J.~Chem.~Phys \textbf{104}, 6976 (1996)]. At
the threshold the photodetachment cross-section displays the
Gailitis--Damburg oscillations.
\end{abstract}
 \maketitle

Considerable attention is payed presently to so-called
dipole-bound anions (DBA), \textit{i. e.} molecular negative ions
in which the excess electron is bound to the molecular neutral~(MN) due
to its dipole moment~\cite{01Compton-Hammer}. Such structures play
important role in various physical, chemical and biological
processes~\cite{91Sonntag}.

As early as in 1947 Fermi and Teller~\cite{47Fermi-Teller} in
their analysis of meson capture by hydrogen atoms noted that a
fixed point dipole~$d>1.625$~D can bind an electron to an
infinitely large number of bound states. A number of subsequent studies
taking into account the finite dipole
effects~\cite{67Levy-Leblond}, presence of a short-range
repulsive core potential~\cite{67Crawford}, rotational
excitation~\cite{Garrett}, polarization and quadrupolar
interaction~(see Ref.~\cite{98Abdoul-Desfrancois} and references
wherein)  yield the minimum dipole moment 2--2.5~D required to
form DBAs of common
molecules~\cite{01Compton-Hammer,94Desfrancois-Abdoul-Khelifa-Schermann}.

DBAs for experimental studies are created mainly by free
electron attachment under high pressure nozzle expansion
conditions~\cite{97Hendricks-deClercq-Lyapustina-Bowen,98Han-Kim-Song-Kim,00Lee-Arnold-Eaton-Bowen,%
02Gutowski-Hall-Adamowicz-Hendricks-deClercq-Lyapustina-Nilles-Xu-Bowen}
and by charge transfer from Rydberg atoms~\cite{94Desfrancois-Abdoul-Khelifa-Schermann,%
96Compton-Carman-Desfrancois-Abdoul-Schermann-Hendricks-Lyapustina-Bowen}.
The created DBAs are investigated, \textit{e.~g.}, by
photoelectron
spectroscopy~\cite{97Hendricks-deClercq-Lyapustina-Bowen,98Han-Kim-Song-Kim,%
00Lee-Arnold-Eaton-Bowen,
02Gutowski-Hall-Adamowicz-Hendricks-deClercq-Lyapustina-Nilles-Xu-Bowen,
96Compton-Carman-Desfrancois-Abdoul-Schermann-Hendricks-Lyapustina-Bowen,
96Bailey-Dessent-Johnson-Bowen}.

Large-scale \textit{ab initio} calculations~(see, for
instance,~\cite{02Gutowski-Hall-Adamowicz-Hendricks-deClercq-Lyapustina-Nilles-Xu-Bowen,%
98Gutowski-Jordan-Skurski,02Peterson-Gutowski,02Sawicka-Skurski}
and references therein) of DBA structure studied the effects of
correlation, orbital relaxation, dispersion and charge-transfer
interaction. However, despite the increasing accuracy of such
many-electron calculations, simplified one-electron local model
potentials still have been proving their efficiency and
suitability for large-scale computer simulations, as well as
more analytical theories~(\cite{01Wang-Jordan} and references
wherein).

In this Letter we use simple model which considers DBA as an
electron moving in a point dipole potential. The developed below
simple analytic theory allows to explain some features of
frequency-dependent DBA photodetachment. While the most of
experimental photoelectron spectra are recorded at a constant
photon frequency~$\omega$, the
Reference~\cite{96Bailey-Dessent-Johnson-Bowen} reports the
photodetachment rate $\propto\omega^{-2}$. Such a behavior is
different from the photodetachment cross-section
$\propto\omega^{-7/2}$ in
atoms~\cite{00Amusia-Avdonina-Drukarev-Manson-Pratt} and
$\propto\omega^{-3/2}$ in atomic negative
ions~\cite{78Demkov-Ostrovsky}, and can be easily explained
using the proposed analytical technique. Our model also predicts
oscillatory behavior of the cross-section near the threshold.
Such oscillations were predicted for electron scattered on
hydrogen atom~\cite{63Gailitis-Dambourg-JETP} and two-photon
photodetachment of hydrogen anion with excitation of the
residual atom~\cite{91Liu-Du-Starace}. The oscillations (as well
as the linear Stark effect) are due to constant dipole moment of
excited states of nonrelativistic hydrogen atom. Similar
oscillations for electron scattering on polar molecules are
discussed in~\cite{77Fabrikant}.

We consider the MN to be a symmetric top whose rotational state
is determined by its angular momentum~$j$ with the
projection~$j_\zeta$ onto the molecule-fixed $\zeta$-axis
directed along the MN dipole moment~$\mathbf{d}$ which is
considered to be a point one. We also assume that MN remains in
one of its vibrational states. The Hamiltonian of the excess
electron moving in the field of the rotating MN is
\begin{eqnarray}
\label{eq:H}
\hat{H}&=&\hat{H}_{\text{rot}}+T_{\text{e}}+V(r,\cos\vartheta)
=b_\xi\hat{j^2}+(b_\xi-b_\zeta)\hat{j_\zeta}\nonumber\\
&+& \frac{\hbar^2}{2m_{\text{e}}r^2}\frac{\rmd}{\rmd r}
\left(r^2\frac{\rmd}{\rmd r}\right)+
\frac{\hbar^2\hat{\mathbf{l}^2}}{2m_{\text{e}}r^2}
-\frac{\beta\cos\vartheta}{r^2},
\end{eqnarray}
where $\mathbf{l}$ is the electron orbital momentum, $b_\xi$
and~$b_\zeta$ are the MN rotational constants, $\mathbf{r}$ is
the electron radius-vector whose direction is determined by the
spherical angles $(\vartheta,\varphi)$ in the molecule-fixed
frame. Dimensionless dipole
moment~$\beta=2m_{\text{e}}|e|d/\hbar^2=0.786~d$(D), $e$ and
$m_{\text{e}}$ being the electron charge and mass.

The solution of Schr\"odinger equation with the
Hamiltonan~(\ref{eq:H}) can be found as a sum over the channels
corresponding to different MN states:
\begin{equation}\label{eq:Psi}
\Psi=\sum_{j}R_{j}(r)\Phi_{j},
\end{equation}
where the angular functions~$\Phi_{j}$ depend on $\vartheta$,
$\varphi$ and on the MN coordinates. There are two limiting cases
when the Schr\"odinger equation with the
Hamiltonian~(\ref{eq:H}) is separable. The first case,
Born--Oppenheimer approximation (BOA), takes place when the MN
rotation is slow compared to the electron motion.
Quantitatively, the difference between the energy levels of the
MN states with neighboring $j$~values in BOA is small compared
with the difference between the excess electron levels. The
opposite limiting case, inverse Born--Oppenheimer approximation
(IBOA) is fulfilled when the electron motion can be considered
to be slow compared to the MN rotation. Thus in IBOA only one
term remains in~(\ref{eq:Psi}) corresponding to the
conserving MN angular moment~$j$; such wavefunctions were used
in rotationally adiabatic theory of rotational autodetachment of
DBA~\cite{88Clary}.

The procedure of the separation of variables is given in detail
in~\cite{92Zon,94Watson} (BOA) and~\cite{95-97Zon} (IBOA). In
both cases the expression~(\ref{eq:Psi}) is effectively reduced
to a single-channel radial function multiplied by the
appropriate angular function. The latter has more simple form in
BOA case, so all the following results (which are not sensitive
to the angular dependence of the electron wavefunction) will be
formulated for BOA case.

The BOA angular
functions~$\mathcal{Z}\equiv\mathcal{Z}_{\lambda\eta}$ satisfy
the equation
\begin{equation}\label{eq:AngEq}
(-\hat{\mathbf{l}^2}+\beta\cos\vartheta)\mathcal{Z}_{\lambda\eta}
=\eta\mathcal{Z}_{\lambda\eta}
\end{equation}
and can be decomposed over the familiar spherical harmonics:
\begin{displaymath}
\mathcal{Z}_{\lambda\eta}(\vartheta,\varphi)=\sum_{l}a_{\eta
l}^{(\lambda)} Y_{l\lambda}(\vartheta,\varphi).
\end{displaymath}
Here the $\zeta$-projection~$\lambda$ of the electron orbital
momentum is conserved due the axial symmetry. This non-spheric
symmetry results in $l$-mixing taken into account by the
coefficients $a_{\eta l}^{(\lambda)}$. Together with these
coefficients, $\mathcal{Z}_{\lambda\eta}$ depend on the
eigenvalue~$\eta$ of the operator~(\ref{eq:AngEq}), and $\eta\to
l(l+1)$, $l=0,1,\ldots$ for $d\to0$. This three-diagonal
eigenvalue problem was studied in~\cite{92Zon,94Watson} for
application to Rydberg states in polar molecules. The
functions~$\mathcal{Z}$ were used as early as in Debye's works
on the Stark effect on polar molecules~\cite{29Debye}. The same
angular functions were used in analysis of the critical binding
dipole moment in DBA~\cite{67Levy-Leblond} without studying the
radial functions; for other references on usage of $\mathcal{Z}$
functions see~\cite{92Zon}.

The radial functions satisfy
\begin{equation}\label{eq:RadEq}
\frac{1}{r^2}\frac{\rmd}{\rmd r}
 \left(r^2\frac{\rmd R}{\rmd r}\right)-
 \frac{\eta}{r^2}R+\frac{2m_{\text{e}} E}{\hbar^2}R=0.
\end{equation}
The solutions of the Eq.~(\ref{eq:RadEq}) for binding states
($E=-\hbar^2\varkappa^2/(2m_{\text{e}})<0$)
are McDonald functions (since they go to zero when
$r\to\infty$):
\begin{displaymath}
R(r)=\frac{1}{\sqrt{r}}K_{\rho}(\varkappa r),\quad
\rho=\sqrt{\eta+1/4 }.
\end{displaymath}
For real $\rho$ these functions diverge at $r\to 0$:
\begin{eqnarray}\label{eq:McDiverge}
K_{\rho}(\varkappa r)\sim(\varkappa r)^{-\rho},\quad r\to 0,
\end{eqnarray}
so the existence (or absence) of binding electron states for
$\eta>-1/4 $ is determined by the behavior of the MN potential
at small~$r$.

If we consider only the binding due to the MN dipole moment,
then these states arise for $\eta<- 1/4  $ when $\rho$ is
imaginary (it is easy to see that $\eta=-1/4 $ corresponds to
the above cited critical value $d=1.625$~D). Applying a formula
similar to~(\ref{eq:McDiverge}) for imaginary~$\rho$ we see that
the functions~$R(r)$ 
oscillate for $r\to 0$ according to (\ref{eq:McDiverge}) and
$R(r)\sim\exp(-\varkappa r)$ when $r\to\infty$.  Assuming
$s=\rmi|\rho|$ the normalized radial wavefunctions are
\begin{eqnarray}\label{eq:Rbound}
R_{\varkappa s}(r)&=&\left(\frac{2\sinh\pi s} {\pi
s}\right)^{1/2} \frac{\varkappa}{\sqrt{r}} K_{\rmi s}(\varkappa
r),\nonumber\\
s&=&\sqrt{|\eta|- 1/4  }, \quad \eta<- 1/4  .
\end{eqnarray}

Oscillating behavior of wavefunctions (\ref{eq:Rbound}) make them
different significantly from the wavefunctions of an electron in
$\delta$-potential. Interestingly, the
functions~(\ref{eq:Rbound}) satisfy all the necessary boundary
conditions for all $E$. In other words, a \emph{continuous
spectrum of bound states} arises in the field of a point dipole
with $\eta<-1/4 $. In this sense it is an unique case in quantum
mechanics (excluding more trivial cases, \textit{e.~g.},
quasicontinuous spectrum of bound states of an electron in a
macroscopic potential box). The ``falling to the center'' (see
the detailed comments in~\cite{LL}) consists in that the
energies of the bound states are not limited from below and the
wavefunctions corresponding to different energies are not
orthogonal.

Of course, this continuous spectrum of bound states is
interesting only from the mathematical point of view, and
connected with the fact that the Sturm--Liouville
problem~(\ref{eq:RadEq}) has a singularity at $r=0$. Physically,
one should regularize the point dipole potential at small $r$.
It can be achieved, for instance, by considering the MN as an
extended dipole~\cite{67Levy-Leblond}, taking into account some
short-range repulsive core
potential~\cite{67Crawford,01Wang-Jordan}. We use the simplest
regularization model of the nonpenetrating core. The boundary
condition in this model reads as $R_{\varkappa\rho}(r_0)=0$ at
some characteristic value of the MN radius~$r_0$. For $\varkappa
r_0\ll 1$ we can use the formula like (\ref{eq:McDiverge}):
\begin{eqnarray*}\nonumber
&&\hspace{-3ex}K_{\rmi s}(\varkappa r)\simeq\frac{\rmi}{2s}
\left[\Gamma(1-\rmi s)
 \left(\frac{\varkappa r}{2}\right)^{\rmi s}\!\!\!
 -\Gamma(1+\rmi s) \left(\frac{\varkappa r}{2}\right)^{-\rmi s}\right] \nonumber\\
 &&\quad=\frac1s\Im\left[\Gamma(1+\rmi s)\left(\frac{\varkappa r}{2}\right)^{-\rmi s}\right]
\end{eqnarray*}
and obtain the electron energy spectrum:
\begin{eqnarray}\nonumber
&&\sin\left[s\ln(\varkappa r_0/2)-\arg\Gamma(1+\rmi s)\right]=0,
\\\label{eq:Enonpenetr}
&&\hspace{-4ex}E_{\lambda\eta
n}=-\frac{2\hbar^2}{m_{\text{e}}r_0^2} \exp\left\{-\frac{2\pi
n}{s}+\frac 1s\arg\Gamma(1+\rmi s)\right\},
\end{eqnarray}
where $n=1,2,\dots$ is the ``principal'' quantum number and the
dependence on $\eta$ and $\lambda$ is included through~$s$. Note
that a formula similar to~(\ref{eq:Enonpenetr}) describes the
spectrum of excited states of an electron in the finite-size
dipole field~\cite{78Drukarev}. 
The
expression~(\ref{eq:Enonpenetr}) should be applied 
carefully to low-excited DBA states; in the below calculation of
photodetachment rate we do not use this formula.

For free electron states
$E=\hbar^2k^2/(2m_{\text{e}})>0$ and the
radial functions are expressed in terms of the Bessel functions
whose order~$\rho$ may be either real or imaginary:
\begin{eqnarray}\label{eq:Rfree}
R_{k,s}(r)&=&\frac{1}{2\sqrt{r}}[J_{\rmi s}(kr)+J_{-\rmi
s}(kr)], \quad\eta<-1/4,
\nonumber\\
R_{k,\rho}(r)&=&\frac{1}{\sqrt{r}}J_{\rho}(kr),\quad\eta>-1/4.
\end{eqnarray}

\begin{figure}
\includegraphics[width=2.5in]{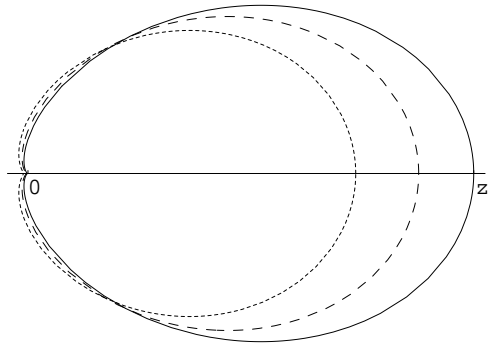}
\caption{Polar directional pattern of
$|\mathcal{Z}_{\lambda\eta}|^2$ for $\lambda=0$ and $d=3$~D
($\eta=-0.74$, dotted line), $d=4$~D ($\eta=-1.18$, dashed
line), $d=5$~D ($\eta=-1.66$, solid line)}
 \label{fig:ang0}
\end{figure}

\begin{figure}
\includegraphics[width=2.5in]{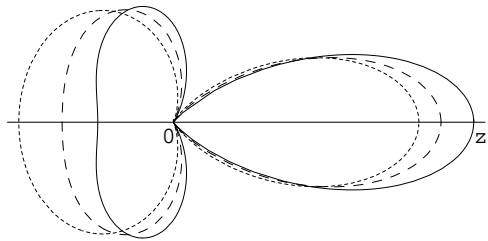}
\caption{Polar directional pattern of
$|\mathcal{Z}_{\lambda\eta}|^2$ for $\lambda=1$ and $d=3$~D
($\eta=2.36$, dotted line), $d=4$~D ($\eta=2.5$, dashed line),
$d=5$~D ($\eta=2.6$, solid line)}
 \label{fig:ang1}
\end{figure}

Figure~\ref{fig:ang0} shows the polar directional patterns of
$|\mathcal{Z}_{\lambda\eta}|^2$ for $\sigma$-states with
$\eta<-1/4 $ at different dipole moment values. Qualitatively,
these patterns resemble those obtained by \textit{ab initio}
calculations~(see, \textit{e.~g.},
\cite{02Gutowski-Hall-Adamowicz-Hendricks-deClercq-Lyapustina-Nilles-Xu-Bowen}).
The same patterns for $\pi$-states with $\eta>-1/4 $ are given
in Fig.~\ref{fig:ang1}

The cross-section of the photodetachment in dipole approximation
is
\begin{displaymath}
\rmd\sigma_{\varepsilon}(\varkappa\lambda\eta;\omega)=\frac{\alpha
m_{\text{e}}k\omega}{2\pi\hbar}
\left|\langle\Psi_{\varkappa\lambda\eta}
|r_{\varepsilon}|\Psi^{(-)}_{\mathbf{k}}\rangle\right|^2d\Omega_{\mathbf{k}}.
\end{displaymath}
Here $\alpha=e^2/\hbar c\simeq 1/137$, $\omega$ is the light
field frequency, $r_{\varepsilon}$ is projection of the electron
coordinate onto the field polarization,
$\Psi_{\varkappa\lambda\eta}$ is the wavefunction of the initial
electron state. The wavefunction~$\Psi^{(-)}_{\mathbf{k}}$ of
the final state has the converging wave asymptotics. Its angular
part can be constructed using the technique developed
in~\cite{92Zon} for BOA and in~\cite{95-97Zon} for IBOA. The
frequency dependence
of~$\rmd\sigma_{\varepsilon}(\varkappa\lambda\eta;\omega)$ is
determined only by the squared radial integral~$Q$ between the
functions~(\ref{eq:Rbound}) and~(\ref{eq:Rfree})
\begin{widetext}
\begin{equation}\label{eq:Hypgeo}
Q\propto\left\{
\begin{array}{ll}
(k/\varkappa)^{\rho'} {}_2F_1(3/2+\rho'/2-\rmi
s/2,3/2+\rho'/2+\rmi s/2;
1+\rho'; -k^2/\varkappa^2),&\eta'>-1/4\\
(k/\varkappa)^{\rmi s'} {}_2F_1(3/2+\rmi(s-s')/2,
3/2+\rmi(s+s')/2; 1+s';-k^2/\varkappa^2)+(s\longrightarrow
s'),&\eta'>-1/4
\end{array} \right.,
\end{equation}
\end{widetext}
where $k^2/\varkappa^2=\hbar\omega/|E|-1$ and the Eq.~7.7(31)
in~\cite{HTF} to rewrite the radial integrals in terms of
Gaussian hypergeometric functions. Depending on the relative
energy of the MN rotation (\textit{i.~e.}, BOA or IBOA), the
parameters~$s$ and $\rho'$ ($s'$) depend on the dipole moment as
well as on~$\lambda$, $\eta$ and the MN rotational quantum
numbers. The final~$\rho'$ ($s'$) depends also on the radiation
polarization~$\varepsilon$.

The full cross
section~$\sigma_{\varepsilon}(\varkappa\lambda\eta;\omega)$ is
the sum over the channels with different $\eta'$, similarly to
the sum over the channels with different orbital momentum $l$ in
spherically symmetric case. The high-frequency limit
$k^2/\varkappa^2\to\infty$ can be obtained
from~(\ref{eq:Hypgeo}) using the asymptotics of the
hypergeometric functions given in~\cite{HTF}, Eq.~2.1(17):
\begin{eqnarray}\label{eq:Highomega}
\sigma(\omega)\propto\omega^{-2},\quad  \hbar\omega\gg |E|.
\end{eqnarray}

The dependence~(\ref{eq:Highomega}) is common for both
perpendicular and parallel ($\varepsilon=x$ or $\varepsilon=z$)
polarization of the radiation as well as for all channels with
$\eta>-1/4 $ and $\eta<-1/4 $. This dependence differs from
photodetachment cross section in $s$-states of atomic negative
ions ($\sigma(\omega)\propto\omega^{-3/2}$) but agrees with the
experimental data~\cite{96Bailey-Dessent-Johnson-Bowen} for DBA.
The difference from the zero-range potential model is due to the
$\propto1/\sqrt{r}$ behavior of the wavefunctions
(\ref{eq:Rfree}) in the small~$r$ domain that is different form
the behavior~$\propto1/r$ of the wavefunctions in 
the zero-range potential. The
$\propto\omega^{-2}$ behavior of $\sigma(\omega)$ holds also
for the non-penetrating core model~(\ref{eq:Enonpenetr})
provided the de~Broglie wavelength of the excess electron is
larger than the effective core radius:
$k\sim\sqrt{2m_{\text{e}}\omega/\hbar}\lesssim 1/r_0$. Assuming
$r_0\sim1$~\AA, the photon frequencies $\sim3$~eV used in
Ref.~\cite{96Bailey-Dessent-Johnson-Bowen} obey this condition.
Actually, for much higher frequencies the ionization of the
inner electronic shells becomes more probable than the
detachment of the dipole-bound electron.

From the Eq.~(\ref{eq:Hypgeo}) one can also extract the behavior
of the photodetachment cross-section at the threshold
$k^2/\varkappa^2\to 0$. Note that the
factor~$(k/\varkappa)^{\rho'}$ ¢ (\ref{eq:Hypgeo}) suppresses
the contribution of the $\eta'>-1/4 $ channels at the threshold,
and the main contribution is given by the $\eta'<-1/4 $
channels. This fact has simple physical explanation. For
$\eta'>-1/4 $, the centrifugal repulsion of the electron exceeds
its attraction by the dipole in the $r\simeq 0$ domain where the
radial wavefunction is localized. Thus the
radial wavefunction is small in $r\simeq 0$ domain because the
electron with $k\to0$ cannot penetrate under the barrier. From
the other hand, the barrier is absent for $\eta'<-1/4 $ so that
the radial function does not disappear in $r\simeq 0$ domain and
the contribution of the corresponding channels remains finite
at~$k\to 0$. So we obtain
\begin{equation}\label{eq:sigmathersh}
\sigma_{x,z}^{(\eta')}(\varkappa\lambda\eta;\omega)\sim
\cos^2\left[s'\ln(k/\varkappa)+\psi_{ss'}\right],
\end{equation}
where the phase~$\psi_{ss'}$ does not depend on~$k$. The
result~(\ref{eq:sigmathersh}) differs significantly from the
power Wigner law of the threshold photoeffect cross-section but
it is similar to the threshold behavior of reactions involving
the excited hydrogen
atom~\cite{63Gailitis-Dambourg-JETP,91Liu-Du-Starace}. Note also
that the dependence~(\ref{eq:sigmathersh}) cannot be obtained in
Born approximation when the dipole influence on the free
electron motion is neglected.

\begin{figure}[h]
\includegraphics[width=3in]{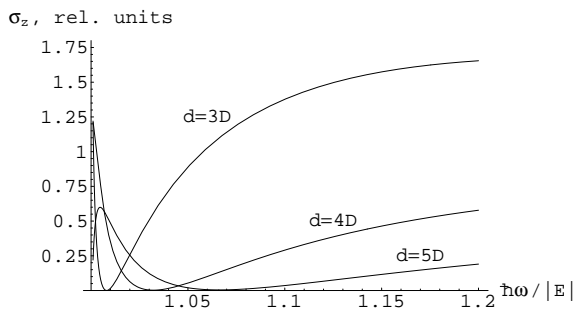}
\caption{Photodetachment cross-section in the near-threshold
domain}
 \label{fig:sigma}
\end{figure}

Figure~\ref{fig:sigma} shows the threshold behavior of the
photodetachment cross-section~$\sigma_z$ when the radiation is
polarized along the dipole direction. For the perpendicular
polarization, the cross-section tends to zero since for
$|\lambda|>0$ one has $\eta>- 1/4  $ (the first value $\eta<-1/4$ 
arises for $\lambda=\pm 1$ at $d\simeq 9.646$~D). The
cross-sections are given in relative units, their absolute
values depend on the approximation (BOA or IBOA) used for the
radial functions. It is seen that the cross-section is
non-monotonic function of $\hbar\omega/|E|$ near the threshold.

We express our gratitude to R.~Compton, pointed out the problem
considered here, to D.~Dorofeev, H.~Helm I.~Kiyan, and S.~I.~Marmo for
helpful discussion. This work was partially supported by CRDF \&
Russian Ministry of Education (award VZ 010-0).

\end{document}